# LearnOpt: Recovering the Latent Cognitive Structure of Standardized Examinations via Knowledge Graphs and Constrained Optimization

**Joy Bose, Om Thomas**
Independent Researchers
Bengaluru, India
joy.bose@ieee.org

**Abstract**

Standardized examinations are typically treated as uniform syllabus coverage problems. We argue they are better understood as adversarial systems with stable latent cognitive structures diverging systematically from official syllabi. We introduce LearnOpt, which recovers this structure from historical question papers and generates personalized, time-bounded study plans. Applied to nine years of NEET questions (2016-2024, n=1,496), LearnOpt builds an exam knowledge graph from LLM-tagged questions, extracts a five-category latent skill distribution, and formulates study planning as a knapsack-variant optimization over prerequisite-aware subgraphs with Bayesian Knowledge Tracing. Central finding: NEET's latent skill distribution is stable within a syllabus regime (consecutive-year KL divergence 0.004-0.032 for 2016-2021, non-significant under permutation testing) but shifts significantly with NCERT's 2023 syllabus rationalization: pooling 2016-2021 (n=1,072) vs 2023-2024 (n=392) gives KL=0.040 (p=0.0005), with Elimination/Negation questions rising from ~20-29% to ~31-35%. Latent structure, while not permanently stationary, is piecewise stable, with shifts detectable and attributable to curricular events. Within either regime, subject predicts skill profile more strongly than year. An optimization evaluation, using one real and two synthetic mastery profiles, shows the skill-weighted objective produces a modest but real reordering of recommended topics over a mastery-conditioned frequency baseline. Applying the pipeline to JEE Advanced reveals a profile dominated by Multi-concept Integration (80.9% vs. 33.3% for NEET), with a JEE-vs-NEET divergence (KL=0.505) exceeding NEET's largest cross-subject divergence: exam tier shapes latent cognitive structure more than subject, which shapes it more than time within a regime. Code, knowledge graph, and annotated dataset are released publicly.

**Plain Language Summary**

*This section explains the paper in simple terms for students, teachers, and coaching centers. The rest of the paper is written for researchers and contains technical details, statistics, and code.*

**What we did.** We took about 1,500 real NEET questions from 2016 to 2024 and asked an AI model to read each one and classify it into one of five types, based on what kind of thinking it actually requires:

- **Just remembering a fact** (e.g., "what is the name of this process")
- **Using an idea in a new situation** (e.g., applying a rule you know to a slightly different example)
- **Connecting two topics together** (e.g., a question that needs both genetics and probability)
- **Doing a calculation** (e.g., working out a number using a formula)

- **Spotting the wrong statement** (e.g., "which of these is NOT correct," "all of the following EXCEPT," or "read these two statements and decide which is true")

**What we found.** Two things stood out.

First, NEET changed in 2023. Before 2023, the mix of these five question types was fairly steady year after year. From 2023 onward, there are clearly fewer pure "remember a fact" questions and clearly more "spot the wrong statement" questions. This lines up with NCERT shortening the syllabus in 2023, fewer chapters seems to mean each remaining topic gets tested more thoroughly, often through these trickier "which one is false" formats. If you are preparing using very old papers (before 2022), the *style* of questions has shifted, even if the topics look similar.

Second, we ran the same check on JEE Advanced and found it looks very different from NEET. JEE Advanced is overwhelmingly "connect two topics together" type questions, with almost no pure recall. NEET is more of a mix. This confirms something coaches already say informally: NEET and JEE reward different kinds of preparation, not just different subjects.

**Why this matters for students.** Most advice says "focus on high-weightage chapters." That is still useful, but it only tells you *what* to study, not *how* the exam will test it. Knowing that NEET now leans more toward "spot the wrong statement" questions means practicing that specific question format (multi-statement questions, Assertion-Reason, Statement I/II type) is worth extra attention, on top of knowing the syllabus.

**What we are building.** Using this analysis, we are building an open-source tool (LearnOpt) where a student can fill in a short form about their own strengths and weaknesses, and the tool suggests a study plan: which topics to prioritize, which to safely de-prioritize, given a fixed amount of time before the exam. This part is still a work in progress, but the question-analysis part above is complete and the underlying data is freely available for anyone to check, including coaching centers who want to verify these patterns against their own question banks.

**For coaching centers.** The tagged dataset (which questions are which type, by year and subject) is published openly. If your own analysis of recent papers shows the same recall-to-elimination shift we describe, that is useful independent confirmation. If it does not, that is useful too: it would mean our sample (drawn from a public online archive, not official NTA papers directly) may not perfectly represent the full exam, which we discuss honestly in the limitations section.

## 1. Introduction

Every year approximately 2.4 million students in India attempt NEET, competing for roughly 100,000 medical seats. The examination covers Biology (90 questions), Chemistry (45 questions), and Physics (45 questions) across a syllabus nominally aligned to NCERT textbooks for Classes 11 and 12. The standard preparation strategy is exhaustive: cover the syllabus uniformly, solve previous-year papers, identify weak topics, repeat.

This strategy has two structural problems. First, it ignores the empirical distribution of the exam. NEET does not sample its syllabus uniformly. Certain topics and crucially, certain cognitive operations appear with systematically higher frequency than others. A student who allocates time uniformly across chapters is misallocating a scarce resource. Second, it conflates topic knowledge with cognitive skill. Two questions classified under the same chapter in the official syllabus may demand entirely different cognitive operations, such as direct fact retrieval versus multi-concept integration, and respond differently to different study strategies.

We introduce LearnOpt to address both problems. The core reframing is simple: exam preparation is a constrained optimization problem. The objective is to maximize expected score. The constraint is available preparation time. The decision variables are which topics and skills to study, in what sequence, given a specific student's current knowledge state.

This reframing has one non-trivial precondition: the exam's structure must be sufficiently stable, at least over identifiable periods, to make historical data predictive of future papers. We call this the stationarity assumption. If the skill distribution of NEET varies wildly and unpredictably from year to year, historical optimization is misleading. If it is stable, or stable within detectable regimes, historical data becomes a reliable signal for future preparation, provided regime shifts can be identified.

Our primary empirical contribution is demonstrating that NEET's latent skill distribution is *piecewise* stationary: stable within a syllabus regime, with a sharp and interpretable shift coinciding with a documented curricular event (NCERT's 2023 syllabus rationalization). This finding is non-obvious in two respects. First, while conventional wisdom among NEET aspirants holds that certain topics recur predictably, the stability of the underlying cognitive structure, the distribution of recall, application, integration, calculation, and elimination questions, had not previously been measured. Second, the finding that this structure can shift, and that the shift is both detectable via a simple divergence metric and attributable to a real external cause, suggests that stationarity should be treated as a monitored property of an exam rather than a fixed assumption baked into an optimization system.

The rest of this paper is organized as follows. Section 2 reviews related work. Section 3 describes the LearnOpt framework. Section 4 presents results. Section 5 discusses implications and limitations. Section 6 concludes the paper.

## 2. Related Work

### 2.1 Bayesian Knowledge Tracing and Student Modeling

Bayesian Knowledge Tracing (BKT), introduced by Corbett and Anderson (1994), models student knowledge as a latent binary state $L_t \in \{0,1\}$ updated via a Hidden Markov Model with four interpretable parameters: initial mastery P(L0), transition probability P(T), slip P(S), and guess P(G). BKT's parameters map directly to educational attributes and can be fit via EM even with sparse data, which has sustained its widespread use. Formally, given an observation $O_t$ at step t ($O_t$=1 for a correct response, $O_t$=0 for incorrect), the posterior knowledge state updates as:

$$P(L_t = 1 \mid O_t = 1) = \frac{P(L_{t-1} = 1)(1 - P(S))}{P(L_{t-1} = 1)(1 - P(S)) + (1 - P(L_{t-1} = 1))P(G)}$$

$$P(L_t = 1 \mid O_t = 0) = \frac{P(L_{t-1} = 1)P(S)}{P(L_{t-1} = 1)P(S) + (1 - P(L_{t-1} = 1))(1 - P(G))}$$

The transition update then advances the state to the next step using the learning-rate parameter:

$$P(L_{t+1} = 1) = P(L_t = 1 \mid O_t) + (1 - P(L_t = 1 \mid O_t))\,P(T)$$

Classic BKT assumes skills are independent, which is unrealistic in curricula with explicit prerequisite chains. Constraint-aware extensions encode prerequisite ordering directly into the transition structure: if concept k1 is a prerequisite of k2, a student cannot transition to mastery of k2 without first demonstrating mastery of k1. This reduces data sparsity and stabilizes latent state estimates in densely interconnected concept graphs, a property directly relevant to LearnOpt's prerequisite-constrained subgraph selection.

Deep Knowledge Tracing (DKT) (Piech et al., 2015) replaced explicit skill boundaries with a continuous latent state modeled via LSTMs, yielding superior predictive accuracy at the cost of interpretability and large data requirements. More recent hybrid frameworks combine an LLM acting as a curricular planner with a knowledge-tracing model acting as an evaluator in a generate-and-retrieve loop, where the KT model scores proposed study sequences via simulated mastery gains.

LearnOpt uses standard BKT with prerequisite-aware constraints. The mastery vector is initialized from self-report rather than logged interactions, and the optimization operates over tens to hundreds of concepts, a regime where BKT's interpretability and low data requirements outweigh DKT-style representational advantages, and where LLM-augmented simulation loops are unnecessary given the explicit graph structure already available.

### 2.2 Automated Educational Knowledge Graph Construction

Manual curation of academic knowledge graphs is expensive and scales poorly. Early automated approaches relied on supervised sequence labeling: BERT-based token classification can discover concept nodes from textbooks with moderate accuracy, but automated prerequisite edge extraction, identifying that "concept A requires concept B", remains substantially harder and typically requires significant human correction (Chaudhri et al., 2021).

LLM-based knowledge graph construction (KGC) shifts this from statistical classification to generative extraction of entity-relation triples directly from raw text. Performance scales with model reasoning capability: comparative evaluations show GPT-3.5-Turbo produces graphs with high literal-extraction rates but heavy node abstraction (81.1% of nodes require post-hoc abstraction), while stronger reasoning models (e.g., o4-mini class) produce denser, more balanced structural maps, and GPT-4o-class models achieve high semantic accuracy with minimal abstraction overhead (2.1%). LLMs nonetheless show variability in how they split and abstract nodes across runs, motivating verification layers.

The SAC-KG (Skilled Automatic Constructor for Knowledge Graphs) framework (Chen et al., 2024) addresses this via a three-agent pipeline: a Generator drafts triples using exemplars from open knowledge bases, a Verifier checks each triple against a RuleHub of 7,000+ structural rules and iterates with the Generator on failure, and a Pruner (a fine-tuned classifier) decides whether tail entities should grow or be pruned. SAC-KG constructs graphs exceeding one million nodes at 89.32% precision, demonstrating that unsupervised, iterative LLM pipelines can approach the precision of manually curated ontologies at a fraction of the cost.

LearnOpt's tagging pipeline (Section 3.2) is considerably smaller in scope, on the order of 1,800 questions rather than million-node ontologies, but adopts the core SAC-KG principle of pairing generation with verification: our chain-of-evidence prompt design functions as a lightweight verifier, requiring the model to justify each skill assignment with a specific textual span, which supports human review of low-confidence labels (Section 4.2).

### 2.3 Learning Path Optimization

Personalized learning path (PLP) generation is standardly framed as finding an optimal traversal of a directed prerequisite graph G = (V,E) from a fictitious source node representing the student's current competency to a target node representing exam requirements. This is formulated as a multi-dimensional Knapsack Problem or Binary Integer Program: select $x_i \in \{0,1\}$ for each concept i to maximize cumulative reward $R_i$ subject to a time budget T and prerequisite ordering constraints, with cost $C_i(m_i)$ a monotonically decreasing function of current mastery m_i. Classical solutions used branch-and-bound, which scales poorly as graph density and constraint count grow. Modern systems instead use CP-SAT solvers (e.g., Google OR-Tools), which convert integer constraints to boolean clauses and use lazy clause generation to find provably optimal solutions for hundreds of variables in milliseconds, enabling real-time replanning as the mastery vector updates, which LearnOpt requires after every study session.

The most directly relevant commercial precedent is Embibe's Personalised Achievement Journey (PAJ), protected under US Patent 11,416,558 (US20200311152A1). The patent describes a system that (1) identifies a subset of a contextualized knowledge graph based on a learner's academic context, (2) filters candidate learning content using behavioral characteristics and historical attempt logs, (3) identifies an optimal path using a concept scoring model and learning strategy, and (4) recommends the resulting personalized path. PAJ operates over a manually curated graph exceeding 75,000 concept nodes and an Embibe Score Quotient (ESQ) combining academic, behavioral, and test-taking features.

LearnOpt adopts PAJ's mathematical skeleton: value/cost evaluation over prerequisite-respecting subgraphs, BKT-based mastery updates, per-session recalibration, via CP-SAT rather than custom dynamic programming heuristics, while replacing the 75,000-node manually curated graph and ESQ behavioral model with LLM-automated graph construction from public exam papers and self-reported mastery. The substantive addition beyond PAJ's architecture is the latent skill extraction layer (Section 3.4): PAJ's concept scoring is defined by curriculum structure, whereas LearnOpt's reward function is additionally weighted by the empirical cognitive-operation distribution of the exam itself (Section 3.5).

### 2.4 Cognitive Taxonomy Calibration and Assessment Analytics

Bloom's Taxonomy provides the standard framework for classifying learning objectives across six levels (Remember, Understand, Apply, Analyze, Evaluate, Create), but six-level schemes show poor inter-rater reliability (IRR) on science MCQs, since raters routinely disagree on the boundary between "remembering" and "understanding" a principle. This is not merely an MCQ artifact: psychometric analysis of two decades of state-level exams using Webb's Depth of Knowledge coding (Webb, 1997) finds that separating higher- from lower-order skills from response data alone is difficult, because items are typically calibrated to a single dominant unidimensional ability, meaning cognitive dimensions must be defined from the surface structure of items, not recovered post-hoc from score matrices.

Collapsed taxonomies substantially improve IRR. In medical education, evaluating student-generated clinical MCQs against a six-level Bloom rubric yielded ICC = 0.54; collapsing to three levels (Recall/Comprehension, Application, Analysis/Evaluation) raised ICC to 0.94 (Grainger et al., 2018). Separately, case-based clinical questions were found to require higher-order cognitive skills 98.1% of the time versus 33.7% for non-case-based questions (Cecilio-Fernandes et al., 2018), indicating that question format, not just content, is a strong predictor of cognitive demand. This is the direct precedent for our S5 (Elimination/Negation) category: format-driven cognitive load is measurable and distinct from topic-driven difficulty.

These findings motivate LearnOpt's five-category taxonomy (Section 3.4): rather than adopting six-level Bloom's or relying on post-hoc statistical separation (shown above to be unreliable for unidimensional exams), we define a small set of categories chosen for surface-structural discriminability, in the spirit of the collapsed three-level rubrics that achieved ICC = 0.94.

On automated assessment assembly, classical Item Response Theory (IRT) under the 2-parameter logistic model characterizes each item by difficulty ($\beta$) and discrimination ($\alpha$) to estimate latent ability $\theta$. Dhavala, Bhatia, Bose, Faldu, and Avasthi (EDM 2020) proposed an automated diagnostic test construction framework that fits deep networks to historical interaction logs via binary cross-entropy to select maximally discriminative items, estimating student ability 40% more accurately than baseline exams with an 18% higher performance spread using fewer items; the approach is further described in US Patent 12,125,412. LearnOpt builds on this prior work (one of the present authors is a co-author of the EDM 2020 paper) but addresses a distinct problem: rather than assembling a diagnostic test from a question bank, LearnOpt characterizes the latent cognitive structure of an existing high-stakes exam and uses that characterization to allocate study time.

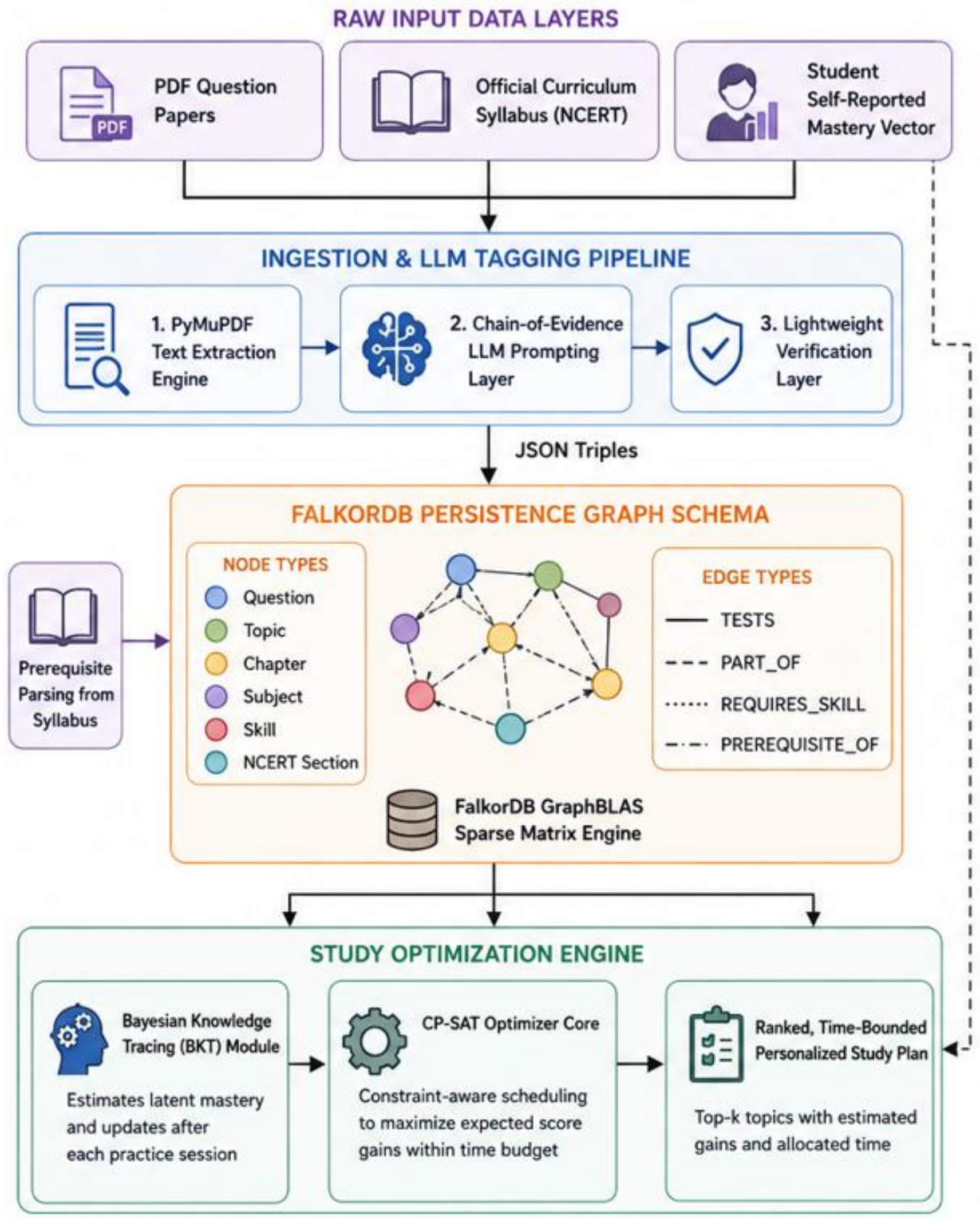

Figure 1: Learnopt System Architecture

## 3. The LearnOpt Framework

### 3.1 Overview

LearnOpt takes as input historical examination papers in PDF format, an official syllabus, and a student mastery vector. It produces a ranked, time-bounded study plan specifying which concept subgroups to study, in what sequence, and with what expected score gain. The framework has four components: Data Ingestion and Tagging, Exam Knowledge Graph, Latent Skill Analysis, and Study Optimization.

### 3.2 Data Ingestion and LLM Tagging

Raw PDF papers are processed using PyMuPDF for text extraction. Each question is extracted into a structured JSON record:

{

"question_id": "NEET_2023_B_045",
"year": 2023,
"subject": "Biology",
"question": "...",
"options": ["A. ...", "B. ...", "C. ...", "D. ..."],
"answer": "B"
}

Each question is then passed to an LLM tagging prompt that extracts:

{
"chapter": "Genetics and Evolution",
"topic": "Pedigree Analysis",
"difficulty": 3,
"skill": "Multi-concept Integration",
"skill_evidence": "Requires combining dominance patterns with probability",
"ncert_reference": "Class 12 Biology Chapter 5"
}

The prompt uses a chain-of-evidence design (Wei et al., 2022) requiring the model to cite the specific reasoning driving the skill classification. This forces justified rather than heuristic tagging and enables human review of low-confidence labels. Tagging reliability is assessed via inter-model agreement (Section 4.2) and a small pilot human annotation by co-author O.T., a practicing NEET aspirant (Section 4.4.1); a full human-annotated gold standard remains future work.

### 3.3 Exam Knowledge Graph

The knowledge graph G = (V, E) is stored in FalkorDB, chosen for its GraphBLAS-backed sparse matrix representation which minimizes multi-hop traversal latency, a practical requirement since the optimization re-runs after each study session and requires fast prerequisite chain traversal.

Node types: Question, Topic, Chapter, Subject, Skill, NCERTSection

Edge types:

- QUESTION -[TESTS]-> TOPIC
- TOPIC -[PART_OF]-> CHAPTER
- CHAPTER -[PART_OF]-> SUBJECT
- QUESTION -[REQUIRES_SKILL]-> SKILL
- TOPIC -[TAUGHT_IN]-> NCERTSection
- TOPIC -[PREREQUISITE_OF]-> TOPIC

Prerequisite edges are inferred via NCERT chapter sequencing and LLM-inferred concept dependencies. Graph schema, ingestion code, and Cypher queries are released in the public repository.

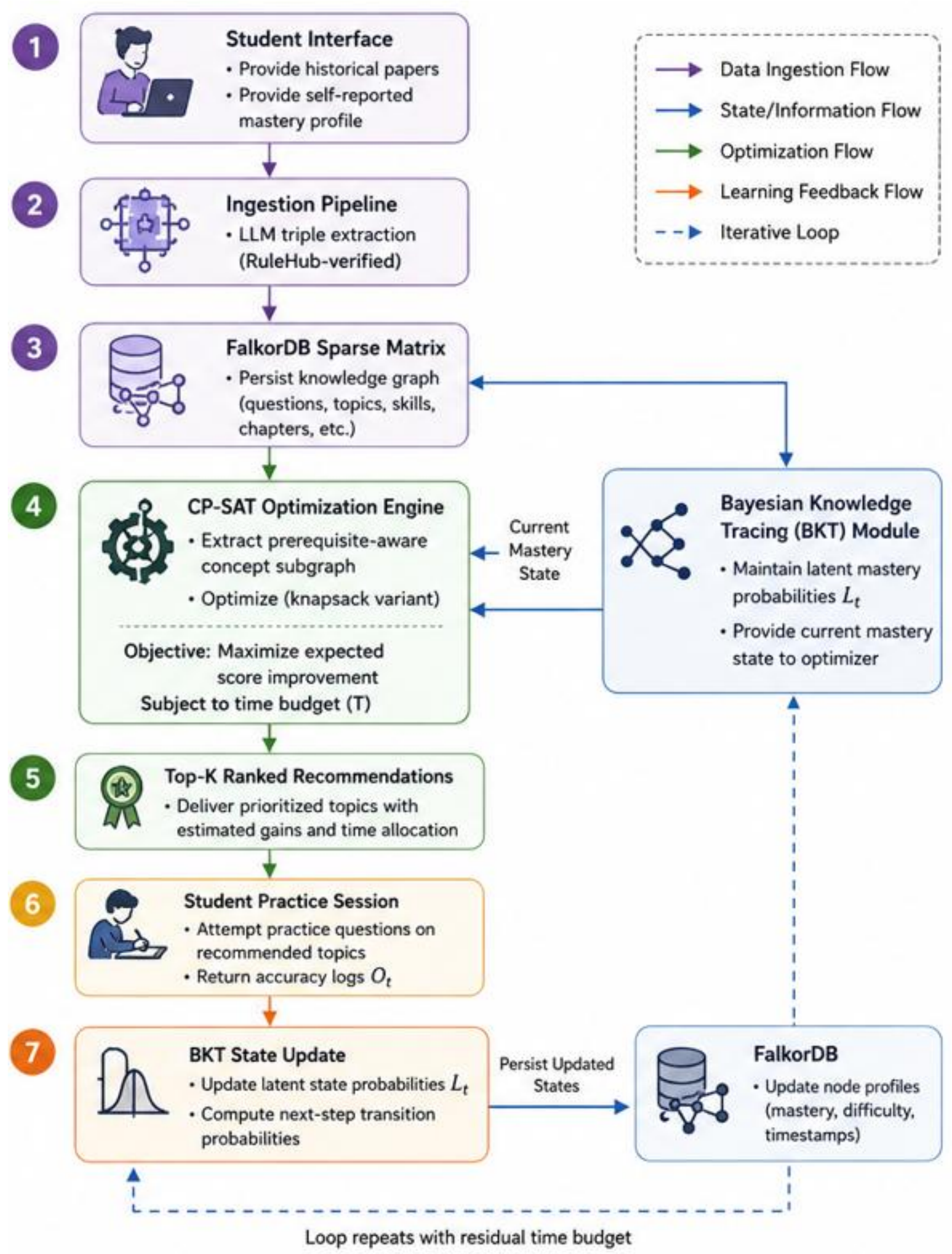

Figure 2: Learnopt Pipeline for Real Time Study Plan Recalibration

### 3.4 Latent Skill Taxonomy

We define five skill categories calibrated to NEET-type science MCQ questions. Categories are chosen for structural discriminability: each can be identified from the surface form of the question without requiring inference about latent cognitive depth, which maximizes inter-rater reliability.

**S1: Direct Recall.** Answer is a single fact retrievable from one NCERT sentence. No reasoning required. Example: "The process of transfer of genetic information from DNA to RNA is called ___."

**S2: Conceptual Application.** A known principle applied to a new context. One inferential step beyond retrieval. Example: standard Mendelian cross with non-standard allele combinations.

**S3: Multi-concept Integration.** Correct answer requires combining knowledge from two or more distinct chapters simultaneously. Example: blood group genetics requiring both inheritance rules and co-dominance.

**S4: Quantitative Reasoning.** Requires numerical calculation, unit analysis, or formula application in a domain context. Example: Hardy-Weinberg allele frequency calculation.

**S5: Elimination and Negation.** Structured as "which is NOT correct" or "all EXCEPT." Requires exhaustive verification of multiple claims rather than retrieval of one answer. This category is analytically distinct from S1-S4: it tests the same content but under a different cognitive operation (falsification rather than retrieval or inference). Its frequency in NEET is a test-design finding rather than a content finding. The precedent for treating question format as a distinct cognitive dimension comes from medical education research, where case-based (format-driven) questions were 98.1% likely to require higher-order cognitive skills versus 33.7% for non-case-based questions on the same content (Cecilio-Fernandes et al., 2018); format predicts cognitive demand independently of topic.

Categories S1-S4 map approximately to the Remember, Understand, Apply, and Analyze levels of Anderson and Krathwohl's revised Bloom's Taxonomy (Anderson & Krathwohl, 2001). S5 does not appear in standard taxonomies and is introduced here to capture a structurally important feature of MCQ design. Our five-category design follows the precedent set by collapsed Bloom rubrics in medical education: a six-level taxonomy applied to clinical MCQs achieved only ICC = 0.54, while a collapsed three-level rubric (Recall/Comprehension, Application, Analysis/Evaluation) achieved ICC = 0.94. LearnOpt's five categories extend this collapsed approach by isolating S5 as a structurally distinguishable sixth dimension, chosen because it can be identified from question format alone, consistent with the broader finding that for unidimensional exams, cognitive dimensions must be defined from item surface structure rather than recovered from response-score matrices.

### 3.5 Study Optimization Engine

**Notation.** Let I be the set of concepts in the knowledge graph. For concept i, let $R_i \in [0,1]$ be its normalized historical frequency (exam relevance weight), $m_i \in [0,1]$ its current mastery level, and $D_i$ its intrinsic difficulty coefficient. Let S be a connected concept subgraph respecting prerequisite edges, and $x_S \in \{0,1\}$ a binary decision variable for subgraph selection.

**Value and cost functions.**

For subgraph S:

$$V_S = \sum_{i \in S} R_i \cdot (1 - m_i) \qquad (1)$$

$$C_S = \sum_{i \in S} D_i \cdot f(m_i) \qquad (2)$$

where $f(m_i)$ is a monotonically decreasing function modeling the learning time required to bridge the remaining knowledge gap, defined as:

$$f(m_i) = (1 - m_i)^{\alpha}, \quad \alpha > 0 \qquad (3)$$

with α calibrated from difficulty ratings.

**Skill-weighted objective.** To couple skill into the optimization directly (rather than treating it as post-hoc characterization), we extend the value function with empirical skill weights $W_s$, which is the observed frequency of skill category s in the exam:

$$V_S = \sum_{i \in S} R_i \cdot W_{\text{skill}(i)} \cdot (1 - m_{\text{skill}(i)}) \qquad (4)$$

where $m_{skill(i)}$ is the student's mastery of the cognitive operation required by concept i, distinct from topic mastery. This ensures that a student weak in multi-concept integration receives more integration-heavy practice regardless of topic.

**Optimization problem.**

$$\begin{aligned} \text{maximize} \quad & \sum_S V_S \cdot x_S \\ \text{subject to} \quad & \sum_S C_S \cdot x_S \le T \\ & x_{S_A} \le x_{S_B} \quad \forall (S_B, S_A) \in E_{\text{prereq}} \\ & x_S \in \{0, 1\} \quad \forall S \qquad (5) \end{aligned}$$

Solved using Google OR-Tools CP-SAT solver. Solution is re-run after each study session as the mastery vector updates via standard BKT.

## 4. Experiments

### 4.1 Dataset

We use 1,496 English-language NEET questions spanning 2016-2024, sourced from a publicly available multilingual NEET question archive (Biology: 743, Chemistry: 375, Physics: 378; per-year n ranges from 32 in 2022 to 200 in 2021, see Table 1). Each question includes the original options and correct answer. NCERT textbooks for Classes 11-12 across Biology, Chemistry, and Physics (both pre- and post-2023 rationalization editions) are used for prerequisite inference and NCERT reference tagging, and to ground the mechanism proposed in Section 4.3 for the 2023 distributional shift.

We note 2022 is substantially underrepresented (n=32 vs. ~170-200 for adjacent years) in this source corpus; we treat 2022 as a transition year of uncertain distribution throughout (Section 4.3) rather than excluding it outright. Extension to the full official NEET corpus (180 questions/year, Biology 90/Chemistry 45/Physics 45) via NTA archives is noted as future work (Section 5.4).

### 4.2 Tagging Reliability: Inter-Model Agreement

As a reliability check on the LLM tagging pipeline, we compute Cohen's kappa (Cohen, 1960) between two independently-prompted LLMs on a common subset of 74 questions that were tagged by both qwen2.5:7b (via local Ollama) and llama-3.1-8b-instant (via Groq, during the retag described in Section 4.4), using the identical chain-of-evidence prompt (Appendix A) at temperature 0.

**Table 1: Inter-model agreement (qwen2.5:7b vs llama-3.1-8b-instant, n=74)**

| Metric | Value |
|---|---|
| Cohen's kappa | 0.510 |
| Raw agreement | 67.6% (50/74) |

A kappa of 0.51 falls in the "moderate" range (Landis & Koch, 1977). The confusion matrix shows this is not uniform noise but a structured pattern: agreement on S5 (Elimination/Negation) is strong (32/40 questions Qwen labeled S5, i.e. 80%, were also labeled S5 by Llama), while disagreement concentrates almost entirely in the S2/S3/S4 cluster, particularly S2-vs-S3 (9 of 27 total disagreements) and S4-vs-S3 (4 of 27). Qwen never assigned S1 on this subset, while Llama did so once; both models agree S1 is rare in this sample, consistent with the 2023-2024 regime values in Table 2.

This pattern is informative about the taxonomy itself: S5 (format-driven, and the category we argue is most structurally distinct, Section 3.4) is reliably identified by both models, while the boundary between "applying a known principle" (S2), "integrating two concepts" (S3), and "applying a formula" (S4) is where models diverge, plausibly because a single question can often be described under any of these framings depending on which step of the reasoning is emphasized. This is an inter-model reliability check, not a validated human gold standard; a full human-annotated gold standard was planned but not completed for this preprint due to annotator availability, and is a priority for a future revision. Kappa=0.51 is likely a lower bound on the taxonomy's reliability, since the two models differ in both scale (7B/8B) and training data, whereas a single consistently-prompted production model (as used for the 1,496- question corpus in Section 4.3, all tagged by llama-3.1-8b-instant after the retag) would be expected to show higher internal consistency than this cross-model comparison suggests.

The paper's central findings (Sections 4.3, 4.5) do not depend on any individual question's label being correct: they are aggregate, regime-level claims over hundreds of questions, where label noise on individual borderline questions (predominantly within the S2/S3/S4 cluster) tends to wash out rather than bias a regime-vs-regime comparison, provided the noise process itself does not differ across regimes, which the tagger-homogeneity check in Section 4.4 was designed to rule out. Moderate per-question kappa is therefore consistent with a high-confidence aggregate finding such as the regime shift in Section 4.3.

### 4.3 Latent Skill Distribution and the 2023 Structural Break

We compute the skill distribution for each year 2016-2024 (n=1,496 questions total; per-year n ranges from 32 in 2022 to 200 in 2021, see Table 1) and measure stability via per-category range and KL divergence (Kullback & Leibler, 1951) between consecutive years. For two discrete skill distributions P and Q over the five categories S = {S1,...,S5}, KL divergence is defined as:

$$D_{\mathrm{KL}}(P \parallel Q) = \sum_{s \in \mathcal{S}} P(s) \log\left(\frac{P(s)}{Q(s)}\right) \qquad (6)$$

Stability is assessed via a permutation testing methodology (Ernst, 2004): under the null hypothesis that two periods share the same skill distribution, year labels are randomly shuffled across the pooled corpus 10,000 times, and the p-value is the proportion of shuffles producing a KL divergence at least as large as observed.

**Table 2: Skill distribution by year (% of questions per category)**

| Year | n | S1 | S2 | S3 | S4 | S5 |
|---|---|---|---|---|---|---|
| 2016 | 172 | 5.8 | 35.5 | 32.6 | 2.9 | 23.3 |
| 2017 | 173 | 5.2 | 33.5 | 31.8 | 4.0 | 25.4 |

| Year | n | S1 | S2 | S3 | S4 | S5 |
|---|---|---|---|---|---|---|
| 2018 | 174 | 6.3 | 25.3 | 36.8 | 2.3 | 29.3 |
| 2019 | 176 | 6.2 | 32.4 | 30.7 | 4.0 | 26.7 |
| 2020 | 177 | 4.0 | 36.2 | 27.7 | 6.8 | 25.4 |
| 2021 | 200 | 4.0 | 34.5 | 38.0 | 4.0 | 19.5 |
| 2022 | 32 | 3.1 | 40.6 | 40.6 | 3.1 | 12.5 |
| 2023 | 193 | 2.6 | 29.0 | 30.6 | 2.6 | 35.2 |
| 2024 | 199 | 2.0 | 19.6 | 42.2 | 5.5 | 30.7 |

**Table 3: KL divergence between consecutive years**

| Transition | KL divergence | Permutation p-value |
|---|---|---|
| 2016 -> 2017 | 0.0036 | 0.957 |
| 2017 -> 2018 | 0.0245 | — |
| 2018 -> 2019 | 0.0196 | — |
| 2019 -> 2020 | 0.0159 | — |
| 2020 -> 2021 | 0.0320 | — |
| 2021 -> 2022 | 0.0247 | 0.846 |
| 2022 -> 2023 | 0.1344 | 0.123 |
| 2023 -> 2024 | 0.0507 | — |
| 2021 -> 2023 (skipping 2022) | 0.0618 | **0.024** |
| **2016-2021 (pooled) vs 2023-2024 (pooled)** | **0.0403** | **0.0005** |

The two regimes are distinguished by a regime-level test, not by any single noisy year-pair comparison. All skill labels in this section were produced by llama-3.1-8b-instant via Groq (a small subset initially tagged by qwen2.5:7b due to rate limits was retagged for tagger consistency; see Section 4.4). From 2016-2021, the distribution is comparatively stable: consecutive-year KL divergence ranges from 0.004 to 0.032, all non-significant under a 10,000-permutation test (e.g., 2016->2017: p=0.957). The single-year 2022->2023 transition (KL=0.134) is *not* statistically significant (p=0.123), but 2022 has only n=32, and a permutation test against such a small sample has low power regardless of effect size. Two tests that avoid this problem both reach significance: the direct 2021->2023 comparison (skipping 2022 entirely, both n>190) gives KL=0.062, p=0.024; and pooling all pre-shift years (2016-2021, n=1,072) against all post-shift years (2023-2024, n=392) gives KL=0.040, p=0.0005. We treat this pooled regime-vs-regime test as the primary evidence for a structural break, since it has the most statistical power and does not depend on any single year's sample size.

The character of the shift: S1 (Direct Recall) falls from a 2016-2021 range of 4.0-6.3% to 1.5-2.6% in 2023-2024, while S5 (Elimination/Negation) rises from 19.5-29.3% to 30.7-35.2%. 2022 (n=32) sits between the regimes on most categories but is too small to characterize reliably on its own and is treated as a transition year of uncertain distribution.

**A plausible mechanism for this can be: NCERT syllabus rationalization.** This timing coincides with a real, documented external event. NCERT released rationalized Physics, Chemistry, and Biology textbooks with multiple chapters deleted entirely from the Class 11-12 syllabus, printed in 2023, and NTA issued a correspondingly revised NEET syllabus with substantial cuts, particularly in Chemistry, ahead of NEET 2024. A smaller, less-codified 2020-21 reduction discussion (COVID-era) does not appear to have produced a detectable shift in our 2016-2021 data, consistent with reports that its adoption into NEET question-setting was unofficial and inconsistent. The 2023 rationalization, by contrast, removed entire chapters from the textbooks question-setters draw from. Fewer available topics plausibly reduces the pool of isolated-fact items amenable to S1 (Direct Recall), while a more concentrated syllabus may shift question-setters toward testing the remaining material more rigorously via multi-statement verification (S5) and cross-topic synthesis (S3).

This mechanism is a plausible explanation consistent with the timing and direction of the observed shift, not a causally established account; we did not have access to internal NTA question-setting processes.

Cross-subject structure persists across both regimes. Despite the temporal break, the relative ordering of skill categories by subject remains informative. Table 4 shows the aggregate skill distribution by subject across all years.

**Table 4: Skill distribution by subject (% of questions, all years)**

| Subject | S1 | S2 | S3 | S4 | S5 |
|---|---|---|---|---|---|
| Biology | 7.8 | 33.1 | 23.0 | 1.5 | 34.6 |
| Chemistry | 1.1 | 26.9 | 34.9 | 7.7 | 29.3 |
| Physics | 0.8 | 29.9 | 51.9 | 7.1 | 10.3 |

Pairwise KL divergence between subjects (Biology vs Chemistry: 0.158, Biology vs Physics: 0.420, Chemistry vs Physics: 0.150) is substantially larger than within-regime year-to-year divergence (2016-2021: max 0.032; 2023-2024: 0.060). This indicates that, within a given syllabus regime, which subject a question belongs to predicts its cognitive-skill profile more strongly than which year it was drawn from. For example, Physics is dominated by S3 (Multi-concept Integration, 51.9%) and nearly devoid of S5, while Biology is comparatively rich in S5 (34.6%) and S1 (7.8%) relative to the other two subjects.

**Revised framing of the stationarity claim.** The original hypothesis, that NEET's latent skill distribution is stable within 5 percentage points across a full decade, is not supported as stated; observed per-category ranges across 2016-2024 are 4.8 (S1), 20.5 (S2), 15.8 (S3), 4.5 (S4), and 25.8 (S5) percentage points. However, a more precise and arguably more useful claim is supported: the skill distribution is stable *within syllabus regimes* and exhibits a detectable, interpretable shift at regime boundaries. This reframes LearnOpt's stationarity precondition (Section 1) from a static assumption to a *monitorable* one, the pipeline developed here could in principle flag when a new exam year's tagged distribution diverges sharply from the recent historical baseline (e.g., KL divergence exceeding a threshold calibrated from the 2016-2021 in-regime range), signaling that historical optimization weights should be recalibrated on post-shift data only.

### 4.4 Tagging Procedure and Cross-Model Homogeneity Check

Of 1,496 questions, all were successfully tagged with a valid skill label using llama-3.1-8b-instant via the Groq API. 136 questions initially encountered API rate limits during the main run; these were first tagged using qwen2.5:7b via local Ollama as a stopgap, but a chi-square test of independence between tagger model and assigned skill category on this subset was significant (chi2=18.74, p=0.0009 within year 2023), indicating qwen2.5:7b systematically assigned more S5/S4 and less S2 than llama-3.1-8b-instant on the same questions. To avoid confounding the regime-shift analysis (Section 4.3) with a tagger artifact, the rate-limited questions were concentrated in 2023-2024, all 136 were subsequently retagged using llama-3.1-8b-instant after API quota reset, yielding a fully tagger-homogeneous dataset (chi2=0.00, p=1.0 by construction). All results in Section 4.3 use this homogeneous tagging. 52 records (3.5%) contained skill_evidence strings that restate the category definition rather than citing question-specific reasoning; these are flagged but retained, and represent an upper bound on potentially low-confidence tags. The 74 questions tagged by both qwen2.5:7b and llama-3.1-8b-instant during this process double as an inter-model reliability check, reported in Section 4.2.

#### 4.4.1 Pilot Human Annotation: A Framing Discrepancy, Not Just Noise

As a small first step toward the human gold standard noted as future work, co-author O.T. (a NEET aspirant) independently labeled 10 real NEET questions (5 from 2018, 5 from 2023) using a simplified four-category scheme presented without reference to the paper's S1-S5 definitions: *Recall* ("you either know the fact or you don't"), *Apply* (a known idea used in a new situation), *Combine* (requires two topics/chapters together), and *Calculate* (needs arithmetic). This is not a like-for-like comparison with the LLM's S1-S5 labels, but it surfaces something the inter-model kappa (Section 4.2) does not: how a *student*, rather than a model, frames these questions.

The result was striking: O.T. labeled **all 10 questions as Recall**, including all 5 questions the LLM tagged S5 (Elimination/Negation), two "which of the following is NOT/EXCEPT" items, one five-statement identify-the-correct-set item, and two Assertion-Reason / twin-statement items.

This is not the human annotator being "wrong," nor evidence against the S5 category, but a real distinction between two senses of "recall." The LLM's S5 label captures a structural property of the question: it requires checking multiple discrete claims against memory before an answer can be selected, regardless of how well any individual claim is known. O.T.'s "Recall" label captures an experiential property: for a student who has already encountered and drilled this question format, the checking process itself feels automatic, and the bottleneck is simply whether each underlying fact is known, i.e., recall, repeated five times, rather than one act of "elimination" as a distinct cognitive operation. Both framings are valid descriptions of the same question, at different levels (question-design structure vs. lived test-taking experience).

This has a direct, practical implication that is consistent with the rest of the paper's findings rather than contradicting them: if S5-format questions are experienced as a sequence of recall checks, then a student's readiness for the post-2023 regime (where S5 makes up 30-35% of questions, Section 4.3) depends on having a correspondingly larger set of well-drilled facts available for rapid-fire verification, not on learning some separate "elimination skill." This single annotator's 10 labels are a suggestive pilot finding about how the taxonomy maps onto student self-perception, not a validation or refutation of the taxonomy's usefulness for the regime-level analysis in Section 4.3. A full human gold standard remains future work.

When asked to generalize beyond the 10 sampled questions, O.T. offered a blunter summary: "all Bio is recall", describing the subject as a whole, not just this sample, as something where the rate-limiting step is whether the underlying fact is known, regardless of the question's surface format. This supports the framing distinction above rather than adding a separate finding: a well-prepared aspirant's lived experience of NEET Biology may be dominated by the "have I memorized this" axis even for questions whose structure (S2-S5) requires application, integration, calculation, or elimination, because for a sufficiently prepared student each of those operations is itself fast and automatic, leaving recall as the only perceptible bottleneck. The S1-S5 structural taxonomy remains the right level of description for how the exam is constructed (Section 4.3's regime-shift analysis is about the exam's design, not any individual student's experience of it), but it is a useful caution against assuming the taxonomy also describes where students should focus their effort, which is a separate question that Section 4.5's optimizer attempts to address via mastery rather than skill category alone.

Relatedly, O.T. described the 2018 and 2023 sample questions used in this pilot (Q1-Q10) as comparatively easy, and reported that 2024-2026 papers (encountered in ongoing test-prep, outside our 2016-2024 corpus) reflect a further format shift toward harder, less recall-amenable questions. This is consistent with the "all Recall" labelling above: a question that is straightforward to a 2026 aspirant, even in S5 format, may not be representative of the harder end of the current question pool.

Motivated by this report, we obtained 180 questions from the 2025 NEET paper and attempted to extend the quantitative analysis to a third regime. However, the available 2025 source material consisted of brief paraphrased question summaries (46-161 characters, e.g. "Find current through battery in given resistor network...") rather than full question-and-options text as used for 2016-2024 (Section 4.1). Tagging this material with two different models (claude-sonnet-4.6 and llama-3.1-8b-instant) produced a KL divergence of 0.248 between the resulting 2024-to-2025 skill distributions (permutation $p<0.0001$ using llama-3.1-8b-instant's tags), the largest divergence reported anywhere in this paper. However, inter-model agreement on this 2025 material was only 30.6% (chi2=103.9, $p<0.0001$), far below the 51% kappa observed for full-text questions (Section 4.2). With only a terse summary rather than the full question and options, different models appear to tag largely on different (and partly fabricated) inferred content, so we cannot attribute the large 2024-2025 divergence to a real shift in the exam. We do not include 2025 in the regime analysis. Reliable skill tagging with this pipeline requires full question text including options, and a full-text 2025/2026 NEET corpus, once compiled in the same format as Section 4.1, is a natural and now well-motivated extension (Section 5.4).

### 4.5 Optimization Evaluation

We evaluate three baselines/methods across three student profiles to test whether the skill-weighted objective (Section 3.5) produces meaningfully different, personalized study plans.

**Methods:**

- B1: Uniform coverage (equal time per topic)
- B2: Topic frequency only (time proportional to historical question frequency, with W_s = 1 for all skills, the dominant real-world heuristic used by NEET aspirants; equivalent to LearnOpt with --no-skill-weights)
- LearnOpt: Skill-weighted objective (W_s = empirical skill-category frequency, Section 3.5)

**Student profiles:**

- **Profile O.T. (real):** a current NEET aspirant's self-assessed mastery across 18 Biology topics (co-author O.T.; see Author Contributions and student_calibration/filled_examples/om_profile.json). Strong (mastery ≥0.8) in Cell Biology, Human Physiology, and Pedigree Analysis; weak (≤0.5) in Biomolecules/Enzymes (0.2) and Molecular Genetics (0.5); three topics (Ecology-Ecosystem, Microbes, Biotechnology) not yet attempted (treated as unrated).

- **Profile A (synthetic):** weak in Genetics/Cell Biology, strong in Ecology/Physiology, the inverse pattern from Profile O.T.
- **Profile B (synthetic)**: the inverse of Profile A.

The key test: B1 and B2 produce identical study plans for all three profiles (they do not condition on mastery beyond the (1-m) cost term, which all three methods share). LearnOpt's skill-weighted ranking additionally re-orders topics by $W_s$, the empirical frequency of each topic's dominant skill category, meaning two topics with identical mastery and historical frequency can receive different priority if one is dominated by S5 (empirically the largest category in the post-2023 regime, Section 4.3) and the other by S4 (the rarest).

**Table 5: Top 5 recommended topics by method and profile (150-hour budget; topic mapping covers 422/1,496 questions, 28.2%, via the keyword bridge described below)**

| Profile | Method | Top 5 topics |
|---|---|---|
| O.T. (real) | LearnOpt (skill-weighted) | Genetics: Molecular basis, Evolution, Coordination Compounds, Reproduction, Ecology: Population |
| O.T. (real) | B2 (topic-frequency) | Genetics: Molecular basis, Evolution, Reproduction, Ecology: Population, Microbes in human welfare |
| Profile A (synthetic) | LearnOpt (skill-weighted) | Genetics: Molecular basis, Reproduction, Cell biology: Cell division, Evolution, Genetics: Mendelian inheritance |
| Profile A (synthetic) | B2 (topic-frequency) | Genetics: Molecular basis, Reproduction, Cell biology: Cell division, Evolution, Cell biology: Organelles |
| Profile B (synthetic) | LearnOpt (skill-weighted) | Ecology: Population, Reproduction, Evolution, Genetics: Molecular basis, Plant physiology: Respiration |
| Profile B (synthetic) | B2 (topic-frequency) | Ecology: Population, Reproduction, Evolution, Genetics: Molecular basis, Plant physiology: Respiration |

**Skill-weighting effect (LearnOpt vs B2, same profile).** For O.T. and Profile A, skill-weighting changes exactly 1 of the top-5 topics (O.T.: "Coordination Compounds" replaces "Microbes in human welfare"; Profile A: "Genetics: Mendelian inheritance" replaces "Cell biology: Organelles"). For Profile B, the top-5 is identical between methods. This is a modest but real effect at the topic-selection level: skill-weighting does not produce a wholesale reordering, but it does promote topics whose dominant skill category is empirically over-represented in the current regime (Section 4.3) over topics with similar frequency/mastery but a less-weighted skill profile.

**Personalization effect (same method, different profiles).** Comparing LearnOpt's top-5 across profiles: O.T. vs Profile A share 3/5 topics, O.T. vs Profile B share 4/5, and Profile A vs Profile B share 3/5. Recommendations are not identical across profiles, consistent with the system conditioning on individual mastery, but substantial overlap remains, driven by a small set of topics (Genetics: Molecular basis, Reproduction, Evolution) that rank highly under most mastery profiles due to high historical frequency combined with moderate-to-low mastery being common across all three profiles for these particular topics.

These are modest effects, not a dramatic personalization demonstration. Two factors likely understate the true effect size. First, only 28.2% of questions map to one of the ~18 calibration topics (see below); the remaining 71.8% retain their original 956 free-text topic groupings, which cannot be personalized against student-reported mastery and default to a neutral value, diluting the mastery-driven signal. Second, Table 5's raw value totals are not directly comparable across methods. LearnOpt's objective includes the W_s factor (always ≤1), so its totals are mechanically smaller than B2's even when the ranking of topics is similar; the meaningful comparison is the ordering (Table 2), not the absolute totals.

A note on topic-level granularity (v0.1). The optimizer groups the 1,496 tagged questions by topic, a free-text field with 956 distinct values produced by LLM tagging (Section 3.2). To connect a subset of these to the ~18 topics used in student self-assessment (Profile O.T. above), we apply a keyword-based mapping (e.g., any topic/chapter containing "pedigree" maps to "Genetics: Pedigree analysis"); the remaining topics retain their original free-text grouping. This 28.2% coverage is itself a useful baseline number for future work: improving it (e.g., via LLM-based topic clustering against the full ~285-chapter set, Section 3.2) is the most direct way to strengthen the personalization signal reported here, and is the clearest concrete next step for the optimization component of LearnOpt. This is a heuristic bridge between the paper's corpus-level analysis (Section 4.3, which operates correctly on the full 956-topic granularity since it only uses the skill field, not topic) and the student-facing tool. A simpler, complementary per-topic heuristic is also released alongside the full subgraph optimizer (Appendix B): for a topic t with historical question count N_q(t), current mastery m_t, and intrinsic difficulty D_t, a return-on-investment score

$$\mathrm{ROI}_t = \frac{N_q(t) \cdot (1 - m_t)}{D_t} \qquad (7)$$

ranks topics for quick, single-step study recommendations, as implemented in the Cypher query in Appendix B.

### 4.6 Generalizability: JEE Advanced

To test whether the skill-distribution methodology generalizes beyond NEET, we apply the identical tagging pipeline (Section 3.2-3.4) to JEE (Advanced), India's exam for admission to the Indian Institutes of Technology, substantially harder than NEET and aimed at a different cohort (engineering rather than medical aspirants).

**Dataset.** We use the single-correct MCQ subset of JEEBench (Arora et al., 2023), a benchmark of 515 JEE Advanced problems spanning 2016-2023. JEEBench includes four response types (MCQ single-correct, MCQ multiple-correct, Integer, Numeric); only the 110 single-correct MCQs (Mathematics: 53, Chemistry: 30, Physics: 27) are structurally comparable to NEET's single-correct format and were tagged. The remaining types (multiple-correct, integer-answer, numeric-answer) do not have a natural "eliminate the incorrect option" reading and are excluded. Coverage is uneven across years (0-27 questions/year, with 2018 entirely absent from the MCQ subset), which precludes a year-level stationarity analysis analogous to Section 4.3; we therefore report only the aggregate distribution.

**Table 6: Skill distribution: JEE Advanced (MCQ subset, aggregate) vs NEET (all years)**

| Skill | JEE Advanced (n=110) | NEET (n=1,496) |
|---|---|---|
| S1 (Direct Recall) | 0.0 | 4.3 |
| S2 (Conceptual Application) | 8.2 | 30.7 |
| S3 (Multi-concept Integration) | 80.9 | 33.3 |
| S4 (Quantitative Reasoning) | 3.6 | 4.5 |
| S5 (Elimination/Negation) | 7.3 | 27.1 |

KL divergence (JEE Advanced vs NEET, overall) = **0.505**.

A clear exam-tier signal. JEE Advanced's profile is dominated by S3 (80.9%) with effectively zero S1, consistent with its reputation as a multi-step, integration-heavy exam rather than a fact-recall exam. NEET, by contrast, is comparatively balanced across S2/S3/S5 (30.7/33.3/27.1) with a non-trivial S1 component (4.3%). Per-subject comparisons (Table 7) show the same pattern within each subject: JEE Advanced Mathematics is 92.5% S3; JEE Advanced Chemistry and Physics are 63.3% and 77.8% S3 respectively, versus 34.9% and 51.9% for the same subjects in NEET.

**Table 7: Same-subject cross-exam comparison**

| Subject | Skill | JEE Advanced | NEET |
|---|---|---|---|
| Chemistry | S3 | 63.3 | 34.9 |
| Chemistry | S5 | 6.7 | 29.3 |
| Physics | S3 | 77.8 | 51.9 |
| Physics | S5 | 14.8 | 10.3 |

KL divergence: Chemistry (JEE vs NEET) = 0.234; Physics (JEE vs NEET) = 0.267.

**Interpretation:** exam tier dominates subject, which dominates time. This produces a clear ordering of effect sizes across the three axes examined in this paper. The JEE-vs-NEET divergence (0.505) exceeds NEET's largest cross-subject divergence (Biology vs Physics, 0.420, Section 4.3 Table 4), which in turn exceeds NEET's largest within-regime cross-year divergence (0.060, 2023-2024). In other words: which exam a question is from predicts its cognitive-skill profile more strongly than which subject it is from, which in turn predicts it more strongly than which year it is from (within a stable syllabus regime). This ordering is intuitive: exam tier reflects a deliberate institutional design choice about what kind of reasoning to test, subject reflects disciplinary differences in how knowledge is structured, and year-to-year variation within a regime reflects only sampling noise around a fixed design.

For LearnOpt as an optimization framework, this has a direct implication: relevance and skill weights ($R_i$, $W_s$ in Section 3.5) must be computed separately per exam, and very likely should not be transferred even between exams that superficially overlap in subject matter (e.g., NEET Physics and JEE Physics), given a same-subject KL divergence of 0.267, comparable in magnitude to NEET's own cross-subject divergences. This confirms LearnOpt's per-exam tagging pipeline, rather than a universal pre-trained skill model, as the architecturally correct choice.

**Limitations specific to this comparison.** The n=110 JEE Advanced sample is small, drawn from a single existing benchmark rather than a purpose-built corpus, and excludes three of four JEEBench response types by construction. The near-zero S1 for JEE Advanced may partly reflect JEEBench's own curation choices (the benchmark's stated goal was to test "harder" problems, which may selectively exclude trivial-recall items even within the MCQ subset) rather than purely reflecting the live exam's composition. A direct extension using JEE *Main* (the higher-volume, MCQ-native exam more comparable in format to NEET, and the exam actually optimized against by the majority of JEE aspirants) is noted as future work in Section 5.4.

In either case, this cross-exam comparison provides the first generalizability evidence for the LearnOpt pipeline beyond a single exam, supporting the domain-agnostic claim made in Section 5.2.

## 5. Discussion

### 5.1 The Latent Fingerprint Finding and Its Breakpoints

NEET's latent skill distribution is a finding about assessment design rather than study strategy, but it is not a single static blueprint, it is a piecewise one. Within the 2016-2021 regime, NEET had an implicit cognitive profile (S2 and S3 each around 30-38%, S5 around 20-29%, S1 around 4-6%) that held steady year over year despite topic-level variation. This profile was never published by the exam board; it is recoverable only from empirical analysis of historical papers. The 2023 rationalization then produced a new profile, most notably, S1 falling toward 2.0-2.6% and S5 rising toward 30.7-35.2% (pooled regime comparison: KL=0.040, permutation p=0.0005).

The S5 (Elimination and Negation) frequency is itself a test-design signal independent of content. At roughly one-fifth to one-third of questions depending on regime and subject (Biology runs highest at 34.6% aggregate), the exam board has made an implicit, substantial commitment to testing exhaustive verification under time pressure, a structural choice about assessment format, not about Biology, Chemistry, or Physics content per se. That this proportion *increased* after the 2023 rationalization is itself informative: a smaller syllabus appears to have been accompanied by *more*, not less, emphasis on this format, plausibly because fewer remaining topics are each tested more thoroughly via multi-statement items.

The practical implication for LearnOpt is that the "exam fingerprint" is best understood as a versioned object: a fingerprint-per-regime, with regime boundaries detectable via the same KL-divergence machinery used to characterize the fingerprint itself (Section 4.3). A deployed system should recompute its relevance weights R_i and skill weights W_s using only post-shift data once a shift is detected, rather than averaging across a regime change.

**Qualitative corroboration from a current aspirant.** Independent of the question-level annotation discussed in Section 4.4.1, co-author O.T. (a current NEET aspirant) was asked, without being shown the quantitative results above, whether the Assertion-Reason / multi-statement format "felt like" it had become more common in recent papers compared to older ones. The response: "Post 2024, papers have been observed to have more application-based questions", an independent, qualitative observation from inside current test-prep culture that a format shift occurred around the same time as our quantitatively detected regime change, even though the respondent's framing ("application-based") differs somewhat from our "elimination/negation" framing (a difference consistent with the framing discrepancy discussed in Section 4.4.1: what we measure structurally as more S5 may be experienced by students primarily as "more application," since verifying multiple statements often itself requires applying a concept to each one). This is one informal data point alongside the NCERT-rationalization hypothesis, not independent statistical evidence.

### 5.2 Generalizability

Section 4.6 demonstrates LearnOpt's pipeline transfers to a second exam (JEE Advanced) without modification beyond pointing at a different question corpus, confirming the framework is not NEET-specific. The same pipeline applies to any exam for which historical papers are publicly available. For LeetCode-style preparation, the hierarchy maps to Problem -> Pattern -> Skill with company-tag frequency as the relevance signal. For UPSC or GATE, the pipeline requires only a different question corpus and syllabus.

The piecewise-stationarity finding (Section 4.3) and the exam-tier finding (Section 4.6) suggest a general principle: an exam's latent skill distribution should be expected to be stable *within* a fixed combination of (syllabus regime, exam tier), and the pipeline's KL-divergence diagnostic can be used to test this for any new exam before relying on its historical data for optimization. Exams undergoing active redesign, such as JEE's own periodic Main/Advanced format revisions, would be expected to show the same kind of detectable regime shift documented for NEET in Section 4.3, and should be checked rather than assumed stationary.

### 5.3 Relation to Embibe PAJ

LearnOpt adopts the mathematical skeleton of Embibe's PAJ system (US Patent 11,416,558 / US20200311152A1), value/cost evaluation over prerequisite subgraphs, BKT-based mastery updates, per-session recalibration, while replacing proprietary infrastructure (a 75,000+ node manually curated graph and the behavioral Embibe Score Quotient) with publicly reproducible components: an LLM-tagged graph built from public exam papers and a self-reported mastery vector. The substantive extension beyond PAJ's architecture is the skill-weighted objective function (Section 3.5), which couples the empirical cognitive-operation distribution of the exam directly into the reward signal, whereas PAJ's concept scoring is defined purely by curriculum structure.

### 5.4 Limitations

Several limitations should be noted. First, mastery vectors are initialized from self-report, introducing subjective bias. Second, optimization evaluation is simulation-based; real student outcomes are not measured. Third, LLM tagging introduces noise that propagates through the graph and optimizer, the kappa score in Section 4.2 quantifies this but does not eliminate it. Fourth, the framework is validated on MCQ exams; transfer to subjective or coding assessments requires further work. Fifth, the approach optimizes for score, which may not align with deeper learning goals. Sixth, as with any analysis of an evolving institutional process, our findings describe these exams as constituted over the years sampled; future papers may diverge from the reported distributions for reasons unrelated to this work, such as routine curriculum or pattern revisions. We do not claim any reported latent skill distribution is permanent, only that it has been stable within the regime and exam studied. Given ongoing NCERT syllabus rationalization (a further reduction was announced for the 2025-26 academic year), and qualitative aspirant feedback that 2024-2026 papers reflect a further format shift beyond what this corpus captures (Section 4.4.1), the 2023-2024 NEET regime characterized in Section 4.3 should itself be treated as provisional; we recommend recomputing the skill distribution and KL-divergence diagnostic against the most recent 1-2 years of any exam before using LearnOpt's weights operationally, using the pipeline released here. Finally, the JEE Advanced analysis (Section 4.6) is based on a small (n=110), benchmark-derived MCQ subset; a direct extension using JEE Main, the higher-volume, MCQ-native exam most aspirants actually optimize against, using a purpose-built corpus analogous to Section 4.1's NEET dataset is left for future work.

## 6. Conclusion

We have presented LearnOpt, a framework that recovers the latent cognitive structure of standardized examinations from historical question papers and uses it to generate personalized, time-bounded study plans. The central empirical finding is that NEET's latent skill distribution is piecewise stationary: stable within a syllabus regime (2016-2021, then 2023-2024, all consecutive within-regime transitions non-significant under permutation testing), with a statistically significant shift between regimes (KL=0.040, p=0.0005) coinciding with NCERT's 2023 syllabus rationalization, a precondition for optimization that is monitorable rather than merely assumed. A second exam, JEE Advanced, shows a markedly different skill profile (dominated by Multi-concept Integration, 80.9% vs. 33.3% for NEET), with the JEE-vs-NEET divergence exceeding even NEET's largest cross-subject divergence, establishing that exam tier shapes latent cognitive structure more than subject, which in turn shapes it more than time within a stable regime.

The broader implication is that standardized examinations possess recoverable stable cognitive fingerprints that diverge from their official syllabi. This is a finding about assessment design with implications for both students and exam boards.

All code, extracted knowledge graph, question-level skill annotations, and optimization scripts are released at https://github.com/joyboseroy/learnopt.

**Author Contributions**

J.B. designed the framework, built the pipeline, ran all experiments, and wrote the manuscript. O.T., a current NEET aspirant, provided domain expertise on NEET preparation: pilot-annotated 10 real NEET questions (Section 4.4.1), provided a real self-assessed mastery profile used in optimization evaluation (Section 4.5), and gave qualitative assessment of the 2023 regime-shift finding and the practical usefulness of the optimizer's recommendation style (Section 5.1), reviewed for factual accuracy from a current aspirant's perspective.

**Appendix A: LLM Tagging Prompt Template**

Given the following NEET question, classify it using exactly one of these five skill categories:

S1 - Direct Recall: Answer is a single fact from one NCERT sentence.

S2 - Conceptual Application: Known principle applied to new context.

S3 - Multi-concept Integration: Requires combining two or more chapters.

S4 - Quantitative Reasoning: Requires calculation or formula application.

S5 - Elimination and Negation: Structured as NOT/EXCEPT question.

Question: {question_text}

Options: {options}

Respond in JSON with fields: skill (S1-S5), skill_evidence (one sentence explaining your classification), chapter, topic, difficulty (1-5), ncert_reference.

**Appendix B: FalkorDB Schema and Sample Cypher Queries**

```
// Skill distribution by year
MATCH (q:Question)-[:REQUIRES_SKILL]->(s:Skill)
RETURN q.year, s.name, count(q) as frequency
ORDER BY q.year, frequency DESC

// Top ROI topics for a student
MATCH (q:Question)-[:TESTS]->(t:Topic)
WHERE t.mastery < 0.5
WITH t, count(q) as question_count
RETURN t.name,
    (question_count * (1.0 - t.mastery)) / t.difficulty as roi
ORDER BY roi DESC LIMIT 20

// Prerequisite chain for a topic
MATCH path = (t:Topic)-[:PREREQUISITE_OF*]->(target:Topic {name: $topic})
RETURN path
```